\documentclass[11pt]{article}
\usepackage{graphics,cite,amssymb,epsfig,float}
\usepackage[usenames,dvips]{color}
\usepackage{amsmath, mathbbol}
\usepackage{subfigure}

\usepackage{multirow}


\def\lsim{\;\raise0.3ex\hbox{$<$\kern-0.75em\raise-1.1ex\hbox{$\sim$}}\;}
\def\gsim{\;\raise0.3ex\hbox{$>$\kern-0.75em\raise-1.1ex\hbox{$\sim$}}\;}

\def\ben{\begin{enumerate}}  \def\een{\end{enumerate}}
\def\bit{\begin{itemize}}    \def\eit{\end{itemize}}
\def\beq{\begin{equation}}   \def\eeq{\end{equation}}
\def\ba{\begin{array}}       \def\ea{\end{array}}
\def\bea{\begin{eqnarray}}   \def\eea{\end{eqnarray}}

\def \1{\hbox{\small 1\kern-3.6pt\normalsize 1}}

\begin{document}

\renewcommand{\thefootnote}{\fnsymbol{footnote}}
\setcounter{footnote}{0}
\vspace*{-2cm}
\begin{flushright}
LPT Orsay 12-109 \\
PCCF RI 12-07\\

\vspace*{2mm}
\today
\end{flushright}
\begin{center}
\vspace*{10mm}
{\Large\bf 
Tree-level
lepton universality violation in the presence of sterile neutrinos: impact for 
$\pmb{R_K}$ and $\pmb{R_\pi}$ 
} \\
\vspace{1cm}

{\bf A. Abada$^{a}$, D. Das$^{a}$, A.M. Teixeira$^{b}$, A. Vicente$^{a}$ and C. Weiland$^{a}$   }

 \vspace*{.5cm} 
$^{a}$ Laboratoire de Physique Th\'eorique, CNRS -- UMR 8627, \\
Universit\'e de Paris-Sud 11, F-91405 Orsay Cedex, France
\vspace*{.2cm} 

$^{b}$ Laboratoire de Physique Corpusculaire, CNRS/IN2P3 -- UMR 6533,\\ 
Campus des C\'ezeaux, 24 Av. des Landais, F-63171 Aubi\`ere Cedex, France

\end{center}

\vspace*{5mm}
\begin{abstract}
We consider a tree-level enhancement to the violation of lepton
flavour universality in light meson decays arising from modified $W
\ell \nu$ couplings in the standard model minimally
  extended by sterile neutrinos.
Due  to the presence of additional mixings 
between the active (left-handed) neutrinos and the new sterile 
states, the deviation from unitarity of 
the leptonic mixing matrix intervening in charged currents 
might lead to a tree-level 
enhancement of $R_{P} = \Gamma (P \to e \nu) / \Gamma (P
\to \mu \nu)$, with $P=K, \pi$.  We illustrate these enhancements
in the case of the inverse seesaw model, showing that one can 
saturate the current experimental bounds on 
$\Delta r_{K}$ (and $\Delta r_{\pi}$), 
while in agreement with the different
experimental and observational constraints.

\end{abstract}
\vspace*{3mm}
\section{Introduction}\label{Sec:Introduction}
Lepton flavour universality (LFU) is one of the distinctive features
of the Standard Model
of strong and electroweak interactions (SM); hence, any deviation from 
the expected SM theoretical estimates in electroweak precision tests 
will signal the presence of New Physics (NP). 
Here we focus on light meson
($K$ and $\pi$) leptonic decays which, in view of 
the expected experimental precision, have a unique potential to 
probe deviations from the SM regarding lepton universality.

In the SM, the dominant contribution to $\Gamma (P \to \ell \nu)$ ($P=K, \pi$)
arises from $W$ boson mediated exchanges. 
The prediction of each specific decay
is heavily plagued by hadronic matrix element 
uncertainties; however, by considering
the ratios 
\begin{equation}\label{eq:RK:Rpi}
R_K \, \equiv \,\frac{\Gamma (K^+ \to e^+ \nu)}{\Gamma (K^+ \to \mu^+
  \nu)}\,,
\quad \quad
R_\pi \, \equiv \,\frac{\Gamma (\pi^+ \to e^+ \nu)}{\Gamma (\pi^+ \to \mu^+
  \nu)}\,, 
\end{equation}
the hadronic uncertainties cancel
out to a good approximation, so that the SM predictions can be
computed with a high precision. 
In order to compare the experimental bounds with the SM predictions, it
proves convenient to introduce a quantity, $\Delta r_P$, which
parametrizes deviations from the SM expectations, possibly arising from
NP contributions: 
\begin{equation}\label{eq:deltar:P}
R_P^\text{exp} \, = \,R_P^\text{SM} \, (1+\Delta r_P) \quad
\text{or equivalently}
\quad
\Delta r_P \, \equiv \, \frac{R_P^\text{exp}}{R_P^\text{SM}} - 1\,.
\end{equation}
The comparison of theoretical 
analyses~\cite{Cirigliano:2007xi,Finkemeier:1995gi} with
the recent measurements from the NA62 
collaboration~\cite{Goudzovski:2011tc} and with the 
existing measurements on pion leptonic decays~\cite{Czapek:1993kc}
\begin{eqnarray}\label{eq:RK:Rpi:SMvsexp}
& R_K^\text{SM} \, = \, (2.477 \pm 0.001) \, \times 10^{-5}\,,
\quad \quad
& R_K^\text{exp} \, = \,(2.488 \pm 0.010) \, \times 10^{-5}\,,
\\
& R_\pi^\text{SM} \, =\, (1.2354 \pm 0.0002) \, \times 10^{-4}\,, 
\quad \quad
& R_\pi^\text{exp} \, =\, (1.230 \pm 0.004) \, \times 10^{-4}\,
\end{eqnarray}
suggest that observation agrees at $1 \sigma$ level with 
the SM's predictions for 
\begin{equation}\label{eq:deltar:P:value} 
\Delta r_K \, = \, (4 \pm 4 )\, \times\, 10^{-3}\,, 
\quad \quad
 \Delta r_\pi \, = \, (-4 \pm 3 )\, \times\, 10^{-3}\,. 
\end{equation}
The current experimental uncertainty
in $\Delta r_K$ (of around 0.4\%) should be further reduced in the
near future, as one expects to have $\delta R_K / R_K \sim
0.1\%$~\cite{Goudzovski:2012gh}, which can translate into 
measuring deviations $\Delta r_K \, \sim
\mathcal{O}(10^{-3})$.
There are also plans for
a more precise determination of $\Delta
r_\pi$~\cite{Pocanic:2012gt,Malbrunot:2012zz}.  

Whether or not $R_P$ can probe a model of NP naturally
depends on the nature and on the expected size of the 
corresponding contributions to $\Delta r_P$. 
LFU in light meson decays can be violated due to 
(i) a new Lorentz structure in the four-fermion interaction (arising
from the exchange of new fields);
(ii) corrections to the 
SM charged current interaction $W \ell \nu$ vertex. 
The first possibility has been extensively discussed in the
literature, especially in the framework of models with an enlarged
Higgs sector, since in the presence of charged scalar Higgs,  new 
tree-level contributions are expected. However, as in the case of most
of Two Higgs Doublet Models (2HDM)\footnote{
As recently shown in~\cite{Celis:2012dk},
the conclusions can be different in a more generalised
2HDM framework.}, or
supersymmetric (SUSY) extensions of the SM, these new tree-level 
corrections are lepton universal~\cite{Hou:1992sy}.
In SUSY models, higher order non-holomorphic couplings can indeed
provide new contributions to $R_P$~\cite{Masiero:2005wr,Masiero:2008cb,Ellis:2008st,Girrbach:2012km,Fonseca:2012kr}, but in view of current
experimental bounds (collider, $B$-physics and $\tau$-lepton decays),  
one can have at most  $\Delta r_K
\leq 10^{-3}$ in the framework of unconstrained minimal 
SUSY models~\cite{Fonseca:2012kr}.
Corrections to the $W \ell \nu$ vertex (case ii) can also induce
violation of LFU in charged currents. 
Loop level corrections to the latter vertex have been considered and,
as referred to in~\cite{Masiero:2008cb}, new particles (at a scale
$\sim \Lambda_\text{NP}$) can indeed account for such a contribution,
but the effect is of order $(\alpha/4 \pi) \times
(m^2_W/\Lambda^2_\text{NP})$ and  generally well below experimental
sensitivity.

The tree-level corrections to charged current 
interactions, once neutrino oscillations are incorporated into the 
SM, are more interesting. 
In this case, and working in the basis where the charged 
lepton mass matrix is diagonal, the
flavour-conserving term $\propto g \bar{l}_j \gamma^\mu P_L \nu_j
W_\mu^-$ now reads 
\begin{equation}\label{eq:cc-lag}
- \mathcal{L}_{cc} = \frac{g}{\sqrt{2}} U_\nu^{ji} 
\bar{l}_j \gamma^\mu P_L \nu_i  W_\mu^- + \, \text{c.c.}\,,
\end{equation}
where $U_\nu^{ji}$ is a generic leptonic mixing matrix,
$i = 1, \dots, n_\nu$ denoting the physical neutrino states
(not necessarily corresponding to the three left-handed SM states $\equiv \nu_L$) 
and $j = 1, \dots, 3$ the charged lepton flavour. 
In the case of three neutrino generations,  $U_\nu^{ji}$ corresponds
to the unitary PMNS matrix and flavour universality is preserved in
meson decays: since one cannot tag the flavour of the final state neutrino
(missing energy), the meson decay amplitude is proportional to 
$(U_{\nu} U_\nu^\dagger)_{jj} =1$, and thus no new contribution to $R_P$ is
expected. 

In order to account for neutrino masses and mixings, the SM can 
be extended with new neutral sterile fermion states ($n_\nu > 3$). 
In many neutrino mass models there are additional singlet states;
the existence of sterile states is also strongly supported by current 
data from reactor experiments, cosmology, as well as indications 
from large structure formation~\cite{Kusenko:2009up}.

In the presence of sterile states,  the $W \ell \nu$ vertex is
proportional to a rectangular 
$ 3\times n_\nu$ matrix $U_\nu^{ji}$, and the mixing
between the left-handed leptons $\nu_L, \ell_L$ corresponds to a $3 \times 3$ block of
$U_\nu^{ji}$,
\begin{equation}\label{eq:U:eta:PMNS}
U_\text{PMNS} \, \to \, \tilde U_\text{PMNS} \, = \,(\1 - \eta)\, 
U_\text{PMNS}\,.
\end{equation}
The larger the mixing between the active (left-handed) neutrinos and
the new states, the more pronounced the deviations from unitarity of
$\tilde U_\text{PMNS}$, parametrized by the matrix $\eta$~\cite{eta}. The
active-sterile mixings and the departure from unitarity of $\tilde
U_\text{PMNS}$ can be at the source of the violation of LFU 
in different neutrino mass models which introduce sterile
fermionic states (gauge singlets, not necessarily right-handed) to
generate non-zero masses and mixings for the light neutrinos.  If
these new states have very small mixings with the left-handed
neutrinos, then $\tilde U_\text{PMNS} \approx U_\text{PMNS}$ and no
effect is expected regarding LFU violation (for instance in the
case of high-scale type I and III fermionic seesaws). On the other
hand, if the model is such that the singlet states are 
not excessively heavy 
and have large mixings with the active ones, then the deviations
from unitarity (i.e. $\!\eta$) can indeed lead to a (tree-level)
correction of the $W \ell \nu$ vertex. 
The impact of the non-unitarity of the lepton mixing matrix on
leptonic light meson decays was first  investigated
in~\cite{Shrock}, prior to the confirmation of neutrino oscillations.
In this work, we revisit this idea in the light of recent neutrino
data and in view of the present (and future) 
experimental sensitivities to 
$\Delta r_{P}$~\cite{Goudzovski:2011tc,Goudzovski:2012gh}.

Corrections to the  $W \ell \nu$ vertex can arise in several
scenarios with additional (light) singlet states, 
as is the case of $\nu$SM~\cite{Asaka:2005an}, low-scale
type-I seesaw~\cite{Ibarra:2010xw} and the Inverse Seesaw
(ISS)~\cite{Mohapatra:1986bd}, among other possibilities. As we
proceed to discuss, should the masses of the new singlet states ($\nu_s$) be
light enough (lighter than the decaying meson), or if the large mixings 
induce significant unitarity violation\footnote{For other 
phenomenological consequences of non-unitarity in the lepton 
mixing matrix, see Ref.~\cite{Malinsky:2009gw}.}
 (even for $m_{\nu_s} \gg m_P$), then 
one might have sizeable $\Delta r_{P}$, within reach of current experimental
sensitivities. 

In the following section~\ref{Sec:drk} we compute in a
model-independent approach $\Delta r_{P}$ in the presence of
additional fermionic sterile states; we then briefly review 
in Section~\ref{Sec:Constraints}  the most important experimental and
observational constraints on the mass of the additional singlet
states.  In Section~\ref{Sec:drkiss}, we consider the case of the
inverse seesaw model to illustrate the impact of sterile neutrinos on
$\Delta r_{P}$. Our concluding remarks are summarised in 
Section~\ref{Sec:conclusions}. 

\section{$\Delta r_K$ in the presence of sterile neutrinos}\label{Sec:drk}

Let us consider the SM extended by $N_s$ additional sterile states, 
and conduct a general formulation of leptonic light meson decays.
The matrix element for the meson decay $P \to l_j \nu_i$ has the generic form
\begin{equation} \label{eq:mat-elem}
\mathcal{M}_{ij} = \bar{u}_{\nu_i} (\mathcal{A}^{ij} P_R + 
\mathcal{B}^{ij} P_L ) v_{l_j} \,,
\end{equation}
no sum implied over the indices of the outgoing leptons $i,j$. Notice
that now one has $i=1, \dots, 3+N_s$. The
expressions for $\mathcal{A}$ and $\mathcal{B}$ can be read from the effective
hamiltonian (obtained after integrating out the $W$ boson in 
Eq. \eqref{eq:cc-lag}),
\begin{equation}\label{eq:eff-ham}
\mathcal{H}_{cc} = \frac{4 G_F}{\sqrt{2}} V_\text{CKM}^{q q^\prime}  
\left( U_\nu^{ji} \bar{q} \gamma_\mu P_L q^\prime \ \bar{l}_j \gamma^\mu P_L
  \nu_i \right)\,,
\end{equation}
implying that in this framework one has
\begin{align}\label{eq:eff-ham2}
& (\mathcal{A})^{ij} \, =\, (\mathcal{A}^W)^{ij}\, =\, 
- 4 \,G_F \,V_\text{CKM}^{us} \, f_P \,U_\nu^{ji \, *}\,  m_{l_j} \,;\\
& (\mathcal{B})^{ij} \, =\, (\mathcal{B}^W)^{ij}\, =\, 
4 \,G_F \,V_\text{CKM}^{us} \,f_P \,U_\nu^{ji \, *}\,  m_{\nu_i}\,,
\end{align}
where $f_P$ denotes the meson decay constant and $m_{l_j, \nu_i}$ the mass of the outgoing leptons.

The expression for $R_P$ is finally given by
\begin{equation}\label{eq:RPresult}
R_P \,= \,\frac{\sum_i 
F^{i1} G^{i1}}{\sum_k F^{k2} G^{k2}}\,, \quad \text{with}
\end{equation}
\begin{align}
& F^{ij}\,=\, |U_\nu^{ji}|^2
\quad \text{and} \ \
 G^{ij} \,=\, \left[m_P^2 (m_{\nu_i}^2+m_{l_j}^2) - 
 (m_{\nu_i}^2-m_{l_j}^2)^2 \right] \left[ (m_P^2 - m_{l_j}^2 -
  m_{\nu_i}^2)^2 - 4 m_{l_j}^2 m_{\nu_i}^2 \right]^{1/2}\,.
  \label{eq:FG}
\end{align}
The result of Eq.~(\ref{eq:RPresult}) has a straightforward
interpretation: $F^{ij}$ represents the impact of new interactions
(absent in the SM), whereas $G^{ij}$ encodes the mass-dependent factors.
Notice however that all states (charged and neutral fermions) do not
necessarily contribute to $R_P$: this can be seen from inspection
of $G^{ij}$, which must be a positive definite quantity. In
particular, we denote by $N_\text{max}^{(l_j)}$ the $N^\text{th}$ heaviest
neutrino mass eigenstate which is kinematically allowed.

The SM result can be easily recovered from Eq.~(\ref{eq:RPresult}),
in the limit $m_{\nu_i} = 0$ and $U_\nu^{ji} = \delta_{ji}$, 
\begin{equation} \label{eq:RMSM}
R_P^{SM} = \frac{m_e^2}{m_\mu^2}
\frac{(m_P^2-m_e^2)^2}{(m_P^2-m_\mu^2)^2} \,, 
\end{equation}
to which small electromagnetic corrections (accounting for
internal bremsstrahlung and structure-dependent effects) should be
added~\cite{Cirigliano:2007xi}.  

The general expression for $\Delta r_P$ now reads
\begin{equation}\label{eq:deltaRPresult}
\Delta r_P \,= \,\frac{m_\mu^2 (m_P^2 - m_\mu^2)^2}{m_e^2 (m_P^2 - m_e^2)^2}\,
\frac{\operatornamewithlimits{\sum}_{m=1}^{N_\text{max}^{(e)}} 
F^{m1}\, G^{m1}}
{\operatornamewithlimits{\sum}_{n=1}^{N_\text{max}^{(\mu)}} 
F^{m2}\, G^{n2}} -1 \,.
\end{equation}
Thus, depending on the masses of the new states (and their hierarchy) and
most importantly, on their mixings to the active neutrinos, $\Delta r_P$
can considerably deviate from zero. In order to illustrate this, we
consider two regimes: in the first (A), all sterile neutrinos are {\it
lighter} than the decaying meson, but heavier than the active neutrino
states, i.e. $m_\nu^\text{active} \ll 
m_{\nu_{s}} \lesssim m_P$; in the second (B), all
$\nu_{s}$ are {\it heavier} than $m_P$. 
Notice that in case (A), all the mass eigenstates can be kinematically
available and one should sum over all $3+N_s$ states; 
furthermore there is an enhancement to $\Delta r_P$ arising from phase
space factors, see Eq. (\ref{eq:FG}).

We further emphasise that scenarios (A) and (B) are in general
experimentally indistinguishable concerning lepton flavour universality, 
the only exception corresponding to
a very particular regime where the sterile neutrinos are very close in
mass to the decaying pseudoscalar meson\footnote{In such a situation,
 the resulting charged lepton would either be less energetic and
   not pass the experimental kinematical cuts~\cite{Shrock}, 
  or then have a clearly reduced momentum.}.  

\section{Constraints on sterile neutrinos}\label{Sec:Constraints}

There are strong experimental and observational bounds 
on the mass regimes and on the size of the active-sterile mixings 
that must be
satisfied. Firstly, it is clear that present data on
neutrino masses and mixings~\cite{Tortola:2012te} should be accounted for. 
Secondly there are robust laboratory bounds from direct sterile
neutrinos searches \cite{Atre:2009rg,Kusenko:2009up}, since the latter can
be produced in meson decays such as $\pi^\pm \to \mu^\pm \nu$, with
rates dependent on their mixing with the active neutrinos. Negative
searches for monochromatic lines in the muon spectrum can be
translated into bounds for $m_{\nu_s} - \theta_{i \alpha}$ combinations,
where $\theta_{i \alpha}$ parametrizes the active-sterile mixing.
The non-unitarity of the leptonic mixing matrix is also subject to
constraints: the rates for leptonic and hadronic processes 
with final state neutrinos depend on $\sum_i |U_\nu^{ji}|^2$, 
where (as mentioned above) the sum extends over 
all neutrino states kinematically accessible ($i=1,\dots,N_\text{max}$), 
and thus constrain the 
departure from the unitarity limit $\sum_i |U_\nu^{ji}|^2 = 1$. 
Bounds on the non-unitarity parameter $\eta$ (Eq.~(\ref{eq:U:eta:PMNS})), 
were derived using Non-Standard Interactions~\cite{Antusch:2008tz};
although not relevant in case (A), these bounds will be taken into account 
when evaluating scenario (B).

Unless the active-sterile mixings are negligible, the
modified $W \ell \nu$ vertex may also contribute to
lepton flavour violation (LFV) processes
\footnote{LFV is typically dipole dominated when the sterile neutrinos
  are light ($m_{\nu_s} \lesssim 300$ GeV), so that $\mu \to e \gamma$
  is the most constraining LFV observable. For heavier sterile neutrinos,
  other (model-dependent) contributions beyond the dipole might be
  more relevant \cite{Abada:2012cq}.}, with potentially large rates.  
  $\mu \to e \gamma$ decays, searched for by the
MEG experiment~\cite{Adam:2011ch}, are the most stringent 
ones~\footnote{Recently, it has been also noticed that in the framework of a low-scale type 
I seesaw, the expected future sensitivity of $\mu- e$ conversion experiments can also play a relevant r\^ole
 in detecting or constraining sterile neutrino scenarios in the 2 GeV - 1000 TeV mass
range~\cite{Alonso:2012ji}.
}- the
rate induced by sterile neutrinos must satisfy~\cite{Ilakovac:1994kj,Deppisch:2004fa}
\begin{equation}\label{eq:BR:muegamma:sterile}
\text{BR}(\mu \to e \gamma) =  
\frac{\alpha_W^3 s_W^2 m_\mu^5}{256 \pi^2 m_W^4 \Gamma_\mu} |H_{\mu e}|^2
\leq 2.4 \times 10^{-12}\, ,
\end{equation}
where $H_{\mu e}  = \sum_i U_\nu^{2i} U_\nu^{1i \, *} 
G_\gamma ( \frac{m_{\nu,i+3}^2}{m_W^2})$, with $G_\gamma$ 
the loop function and $U_\nu$ the mixing matrix defined in  
Eq. \eqref{eq:cc-lag}.  Similarly, any 
change in the $W \ell \nu$ vertex will also affect other leptonic meson
decays, in particular $B \to \ell \nu$; the following bounds were
enforced in the analysis: $\text{BR}(B \to e \nu) < 9.8 \times 10^{-7}$, 
$\text{BR}(B \to \mu \nu) < 10^{-6}$ and 
$\text{BR}(B \to \tau \nu) = (1.65 \pm 0.34) \times 10^{-4}$~\cite{Beringer:1900zz}.

Important constraints can also be derived from LHC Higgs 
searches~\cite{Dev:2012zg} and electroweak precision 
data~\cite{delAguila:2008pw}. 
LHC data on Higgs decays already provides some important bounds 
when the sterile states are slightly below $125$ 
GeV (due to the potential $H$-decays to left- and right-handed 
neutrinos). The active-sterile mixings can introduce small 
deviations to the electroweak fits, which allows to constrain them.  
An effective approach was 
applied in~\cite{delAguila:2008pw}, assuming  very heavy sterile neutrinos, and 
thus these bounds will only be applied in scenario (B).

Under the assumption of a standard cosmology, the most constraining
bounds on sterile neutrinos stem from a wide variety of cosmological
observations \cite{Smirnov:2006bu,Kusenko:2009up}. 
Using Large Scale Structure (LSS) data, one can also set relevant bounds on 
very light sterile neutrinos ($m_{\nu_s} < 100$ eV), since if such light states 
constitute a non-negligible fraction of the dark matter of
the Universe, then structure formation is affected. 
Active-sterile mixing also induces radiative decays
$\nu_i \to \nu_j \gamma$, well constrained by cosmic X-ray 
searches. Lyman-$\alpha$ limits, the existence of additional degrees
of freedom at the epoch of Big Bang Nucleosynthesis, and Cosmic
Microwave Background (CMB) data, also allow to set additional bounds in the
$m_{\nu_s} - \theta_{i \alpha}$ plane.  However, all the above
cosmological bounds can be evaded if a non-standard cosmology is
considered. In fact, the above 
cosmological constraints 
disappear in scenarios with a low reheating
temperature~\cite{Gelmini:2008fq}. 
In our numerical analysis we will allow for the violation of the latter 
bounds, explicitly stating it.

\section{$\Delta r_K$ in the  inverse seesaw model}\label{Sec:drkiss}

Although the generic idea explored in this work 
applies to any model where the active neutrinos have sizeable
mixings with some additional singlet states,  we  consider the case of the Inverse 
Seesaw~\cite{Mohapatra:1986bd} to illustrate
the potential of a model with  sterile neutrinos regarding 
tree-level contributions to light meson decays.
As mentioned before, there are other
possibilities~\cite{Asaka:2005an,Ibarra:2010xw}.

\subsection{The inverse seesaw model}
In the ISS, the SM particle content is extended by $n_R$
generations of right-handed (RH) neutrinos  $\nu_R$ and $n_X$ generations of
singlet fermions $X$ with lepton number $L=-1$ and $L=+1$,
respectively~\cite{Mohapatra:1986bd} (such that $n_R+n_X = N_s$).
Even if deviation from unitarity can occur for different values of $n_R$ and $n_X$,
here we will consider the case $n_R = n_X = 3$. 
The lagrangian is given by 
\begin{equation}
\label{eq:L_IS}
\mathcal{L}_\text{ISS} = 
\mathcal{L}_{SM} + Y_{\nu}^{ij} \bar{\nu}_{R i} L_j \tilde{H}
+ {M_R}_{ij} \, \bar{\nu}_{R i} X_j + 
\frac{1}{2} {\mu_X}_{ij} \bar{X}^c_i X_j + \, \text{h.c.}
\end{equation}
where $i,j = 1,2,3$ are generation indices and $\tilde{H} = i \sigma_2
H^*$. Notice that the present lepton number assignment, together
with $L=+1$ for the SM lepton doublet, implies that 
the ``Dirac''-type right-handed
neutrino mass term $M_{R_{ij}}$ conserves lepton number, while the
``Majorana'' mass term $\mu_{X_{ij}}$ violates it by two units.

The non-trivial structure of the neutrino Yukawa couplings $Y_\nu$ implies
that the left-handed neutrinos mix with the RH ones after electroweak
symmetry breaking.
In the $\{\nu_L,{\nu^c_R},X\}$ basis, one has the following symmetric 
($9\times9$) mass matrix $\mathcal{M}$,
\begin{eqnarray}
{\cal M}&=&\left(
\begin{array}{ccc}
0 & m^{T}_D & 0 \\
m_D & 0 & M_R \\
0 & M^{T}_R & \mu_X \\
\end{array}\right) \, .
\label{nmssm-matrix}
\end{eqnarray}
Here $m_D= \frac{1}{\sqrt 2} Y_\nu v$, with $v$ the vacuum expectation
value of the SM Higgs boson.  
Assuming $\mu_X \ll m_D \ll M_R$, 
the diagonalization of ${\cal M}$ leads to an effective Majorana
mass matrix for the active (light) 
neutrinos~\cite{GonzalezGarcia:1988rw},
\begin{equation}\label{eq:nu}
m_\nu \simeq {m_D^T M_R^{T}}^{-1} \mu_X M_R^{-1} m_D \, ,
\end{equation}
whereas the remaining 6 sterile states  
have masses approximately given by
$M_{\nu} \simeq M_R$.  

In what follows, and 
without loss of generality, we work in a basis where $M_R$ 
is a diagonal matrix (as are the charged lepton Yukawa
couplings). $Y_\nu$ can be written using a 
modified Casas-Ibarra parametrisation~\cite{Casas:2001sr} 
(thus automatically complying with light neutrino data),
\begin{equation}\label{eq:CI:param}
Y_\nu = \frac{\sqrt{2}}{v} \, V^\dagger \, \sqrt{\hat M} \, R \,
\sqrt{{\hat m}_\nu} \, U_\text{PMNS}^\dagger \, ,
\end{equation}
where $\sqrt{{\hat m}_\nu}$ is a diagonal matrix containing the square roots 
of the three eigenvalues of $m_\nu$ (cf. Eq. \eqref{eq:nu}); likewise
$\sqrt{\hat M}$ is a (diagonal) matrix with the square 
roots of the eigenvalues of $M = M_R \mu_X^{-1} M_R^T$.  $V$ diagonalizes 
$M$ as $V M V^T = \hat{M}$, and $R$ is a $3 \times 3$ complex 
orthogonal matrix, parametrized by $3$ complex angles, 
encoding the remaining degrees of freedom. 

The distinctive feature of
the ISS is that the additional $\mu_X$ parameter 
allows to accommodate the smallness of the active neutrino 
masses $m_\nu$ for a low seesaw scale, but  with
natural Yukawa couplings ($Y_\nu\sim
{\mathcal{O}}(1) $). 
As a consequence, one can have
sizeable mixings between the active
neutrinos and the additional sterile states. This is 
in contrast to the canonical type-I seesaw, where
$\mathcal{O}(1)$ Yukawa couplings require $M_R \sim 10^{15}$ GeV, thus
leading to truly negligible active-sterile mixings.

The nine neutrino mass eigenstates enter the leptonic charged current
through their left-handed component (see Eq. \eqref{eq:cc-lag}, 
with $i = 1, \dots, 9$, $j
= 1, \dots, 3$). The unitary leptonic mixing matrix 
$U_\nu$ is now defined as $U^T_\nu \mathcal{M} U_\nu =
\text{diag}(m_i)$. Notice however that only the rectangular $3 \times 9$
sub-matrix (first three columns of $U_\nu$) appears
in Eq. \eqref{eq:cc-lag} due to the gauge-singlet 
nature of $\nu_R$ and $X$.

In the ISS limit ($\mu_X \ll m_D \ll M_R$), and 
following~\cite{Forero:2011pc}, one can expand the neutrino mass
matrix in powers of $\epsilon \equiv m_D \, M_R^{-1}$,  
block-diagonalizing it at leading order in $\epsilon$, thus easily 
obtaining  $U_\nu$,  
and the relevant active-sterile neutrino mixing angle.

\subsection{Numerical evaluation of $\Delta r_K$ in the inverse seesaw model}
We numerically evaluate the contributions to $R_K$ in the framework  of the ISS 
and address the two scenarios discussed before, which can 
be translated in terms of 
ranges for the (random) entries of the $M_R$ matrix:
{\it scenario (A)} ($m_{\nu_s} < m_P$) - ${M_R}_{i} \in [0.1,200]$ MeV; 
{\it scenario (B)} ($m_{\nu_s} > m_P$) - ${M_R}_{i} \in [1,10^6]$ GeV.
The entries of $\mu_X$ have also been randomly 
varied in the $[0.01$ eV$, 1$ MeV$]$ range for both cases.

The adapted Casas-Ibarra parametrisation for $Y_\nu$, 
Eq.~(\ref{eq:CI:param}),
ensures that neutrino oscillation data 
is satisfied (we use the best-fit values of the global
analysis of~\cite{Tortola:2012te}, and set the  
CP violating phases of $U_\text{PMNS}$ to zero). The $R$ matrix  
angles are taken to be real (thus no contributions 
to lepton electric dipole moments are expected), and randomly varied in 
the range ${\theta}_{i} \in [0,2 \pi]$. 
Although we do not discuss it here, 
we have verified that similar $\Delta r_K$ contributions 
are found when considering the more general complex $R$ matrix case.

\begin{figure}[ht]
\begin{tabular}{cc}
\hspace*{5mm}{\footnotesize Scenario (A)} &
\hspace*{11mm}{\footnotesize Scenario (B)}\vspace*{2mm} \\
\epsfig{file=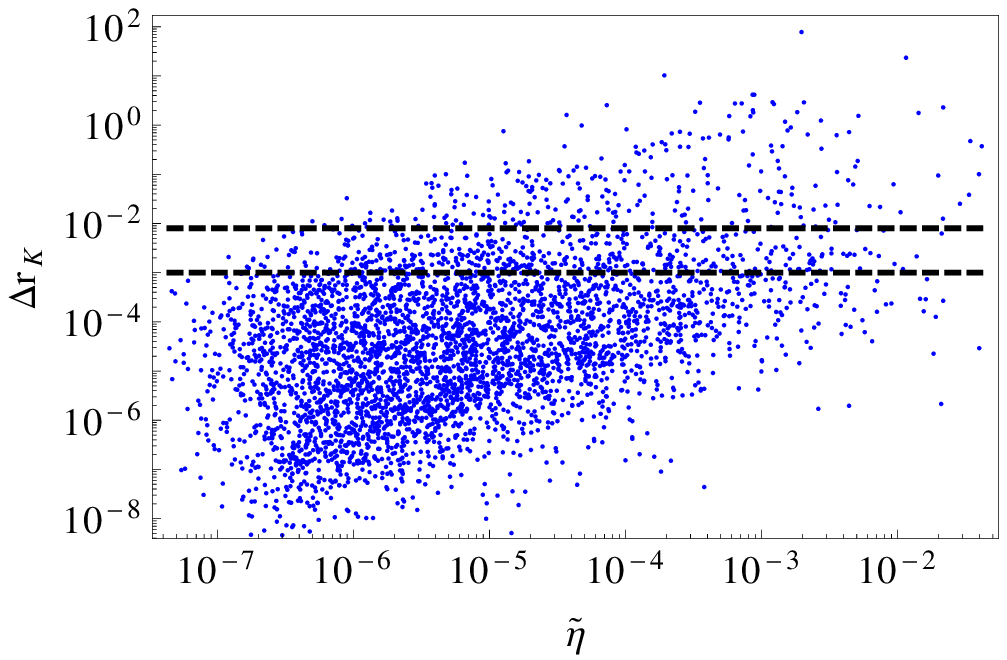, width=75mm} \hspace*{2mm} 
&
\hspace*{2mm}
\epsfig{file=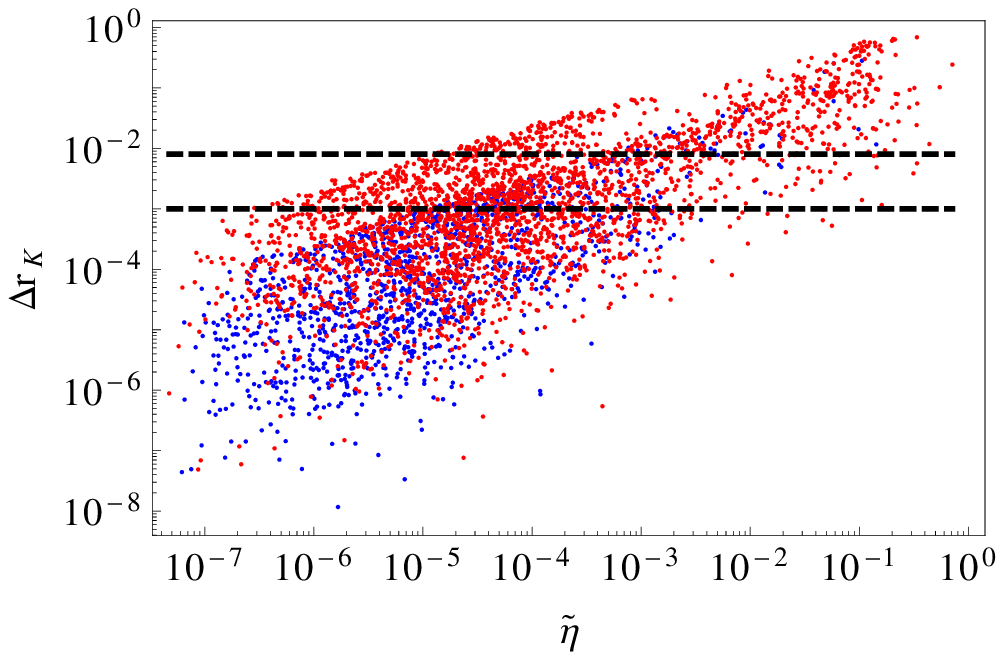, width=75mm} 
\end{tabular}
\caption{Contributions to $\Delta r_K$ in the inverse seesaw as a
  function of $\tilde \eta = 1 - |\text{Det}(\tilde U_\text{PMNS})|$:
  scenarios A (left) and B (right). The upper (lower) dashed line
  denotes the current experimental limit (expected
  sensitivity).  On the right panel, red points denote
    cases where $Y_\nu \gtrsim 10^{-2}$.
 All points  comply with experimental and laboratory constraints. 
 Points in (B) are also in agreement with cosmological bounds, while those in (A) 
 require considering a non-standard cosmology.       } 
\label{figure12}
\end{figure}

In Figs.~\ref{figure12}, we collect our 
results for $\Delta r_K$ in scenarios (A)
- left panel - and (B)
- right panel, as a function of $\tilde \eta$, which parametrizes the 
departure from unitarity of the active neutrino
mixing sub-matrix $\tilde U_\text{PMNS}$, 
$\tilde \eta = 1 - |\text{Det}(\tilde U_\text{PMNS})|$. 
Although the cosmological constraints are not always satisfied, 
we stress that all points displayed comply with 
the different  
experimental and laboratory bounds discussed before.
For the case of scenario (A),
one can have very large contributions to $R_K$, which can even reach
values $\Delta r_K \sim \mathcal{O}(1)$ (in some specific cases we
find $\Delta r_K$ as large as $\sim 100$).  The hierarchy of the
sterile neutrino spectrum in case (A) is such that one can indeed have
a significant amount of LFU violation, while still avoiding
non-unitarity bounds.  Although this
scenario would in principle allow to produce sterile neutrinos in
light meson decays, the smallness of the associated $Y_\nu$
($\lesssim\mathcal{O}(10^{-4})$), together with the loop function
suppression ($G_\gamma$), precludes the observation of LFV processes,
even those with very good  associated experimental sensitivity, as
is the case of $\mu \to e \gamma$.  The strong constraints from CMB
and X-rays would exclude scenario (A); in order to
render it viable, one would require a non-standard cosmology.

Despite the fact that in case (B) the hierarchy of the sterile states
is such that  
non-unitarity bounds become very 
stringent (since the sterile neutrinos are not kinematically viable
meson decay final states),  
sizeable LFU violation is also possible, with deviations from the SM
predictions again as large as $\Delta r_K \sim \mathcal{O}(1)$.  
Contrary to case (A), whose results could also arise in other
frameworks with light sterile neutrinos, 
the large deviations in (B) typically occur when all
 the singlet states are considerably heavier than the decaying meson,
and reflect specific features of the ISS.
As can be inferred from Eq.~(\ref{eq:nu}), in the inverse seesaw
framework, one has 
$m_\nu \sim (Y_\nu \,v / M_R)^2 \, \mu_X$; hence, for ``low'' 
(when compared to, for instance,  the type I seesaw scale) 
$M_R$, light neutrino data can still 
be accommodated with large Yukawa
couplings, $Y_\nu \sim \text{few} \times 10^{-1}$. As a consequence,
large active-sterile mixings can occur, thus leading to an enhancement
of $R_K$.  Even if  in 
this case one cannot produce 
sterile states in meson decays, the large $Y_\nu$ open the
possibility of having larger contributions to LFV observables so that, for
example, BR($\mu \to e \gamma)$ can be within MEG reach in this case.

Although we do not explicitly display it here, the prospects for
$\Delta r_\pi$ are similar: in the same framework, one could have 
$\Delta r_\pi \sim \mathcal{O}(\Delta r_K)$, and thus $\Delta r_\pi \sim \mathcal{O}(1)$ 
in both scenarios. 
Depending on the singlet spectrum, these observables can also be strongly 
correlated: if all the sterile states are either lighter than the pion (as it is the case of scenario (A)) 
or then heavier than the kaon, one finds $\Delta r_\pi \approx \Delta r_K$. 
The latter possibilities are a feature of the ISS mechanism 
(not possible in the unconstrained MSSM, for example) and are  expected to be present 
in other low-scale seesaw models that allow for large active-sterile mixing angles. 

\section{Concluding remarks}\label{Sec:conclusions}
The existence of sterile neutrinos can potentially lead to a significant 
violation of lepton flavour universality 
at tree-level in light meson decays. As shown in this study, 
provided that the active-sterile mixings are sufficiently large, the
modified  
$W \ell \nu$ interaction can lead to large contributions to lepton
flavour universality 
observables, with measurable deviations from the standard model
expectations, well within experimental sensitivity.
This mechanism might take place in a number of frameworks, the 
exact contributions for a given observable being model-dependent. 

As an illustrative (numerical) example, 
we have evaluated the contributions to $R_K$ in the inverse 
seesaw extension of the SM - a truly  minimal extension of the SM - , for
distinct hierarchies of the sterile states. 
In particular, we have studied the 
impact of non-unitarity in a low mass regime for the additional
singlets, an inverse seesaw  mass regime  
considerably lower than what had been 
previously addressed~\cite{Malinsky:2009gw,Antusch:2008tz}.
Recent studies~\cite{Lello:2012gi} have proposed a search of the monochromatic peak in the next generation of  high intensity
experiments,  yielding  both the mass and mixing angles for
sterile neutrinos with masses in the range 
3~MeV $\lesssim m_{\nu_s}   \lesssim 414$~MeV.

Our analysis reveals that very large deviations from the SM
predictions can be found ($\Delta r_K \sim \mathcal{O}(1))$ - or even larger, well
within reach of the NA62 experiment at CERN. 
This is in clear contrast
with other models of new physics (for example unconstrained SUSY
models, where one typically has $\Delta r_K \lesssim
\mathcal{O}(10^{-3})$).  We further notice that these large 
deviations are a generic and non fine-tuned feature of this model.
It is worth emphasising that, in view of the
potentially large new contributions to these observables, such an
analysis of LFU violation in light meson decays actually allows to set
bounds on the amount of unitarity violation (parametrized by $\eta$).

Interestingly, in this framework, both  $\Delta r_K$ 
and $\Delta r_\pi$ are strongly correlated in the case where all the sterile states are lighter than the pion or heavier than the kaon.

The impact of this mechanism is not restricted to light meson decays:
there are currently some hints of lepton flavour universality violation in heavy
mesons, with deviations already found in observables such as
$ \Gamma(B^- \to \tau \nu)/\Gamma(\bar{B}^0 \to \pi^+
\ell^- \nu)$. We expect significant contributions to $B$-meson
observables~\cite{paper2}, which have very promising experimental
perspectives.

\section*{Acknowledgements}
We are grateful to Damir Becirevic, Jorge de Blas and Jean-Pierre Leroy  for
many useful and enlightening discussions. We are also thankful to 
Robert Shrock for valuable exchanges.
This work has been partly done under the ANR project CPV-LFV-LHC
NT09-508531.  The authors acknowledge partial support from the
European Union FP7 ITN INVISIBLES (Marie Curie Actions, PITN- GA-2011-
289442).

\end{document}